\documentclass[runningheads]{llncs}

\usepackage[T1]{fontenc}
\usepackage{graphicx,array}
\usepackage{adjustbox}
\usepackage{array}
\usepackage{booktabs}
\usepackage[inline]{enumitem}
\usepackage{multirow}
\usepackage{multicol}
\usepackage{amssymb}

\newcolumntype{R}[2]{%
    >{\adjustbox{angle=#1,lap=\width-(#2)}\bgroup}%
    l%
    <{\egroup}%
}

\usepackage{tabularx}
\usepackage{makecell}

\usepackage{float}

\usepackage[switch]{lineno}

\usepackage[hyphens]{url} 

\usepackage{textcomp}

\usepackage{algorithm}
\usepackage{algpseudocode}
\usepackage{subcaption}

\usepackage[%
    binary-units=true,
    prefixes-as-symbols=false,
    ]{siunitx}

\AtBeginDocument{%
  \providecommand\BibTeX{{%
    \normalfont B\kern-0.5em{\scshape i\kern-0.25em b}\kern-0.8em\TeX}}}

\usepackage{tikz}

\begin{document}


\DeclareSIUnit{\microsecond}{\SIUnitSymbolMicro s}


%
\title{ Salty Seagull: A VSAT Honeynet to Follow the Bread Crumb of Attacks in Ship Networks
\thanks{
}
}
\titlerunning{}
%
\author{Georgios Michail Makrakis\inst{1}\orcidID{0000-0002-1280-6568} \and
Jeroen Pijpker\inst{1}, \inst{3}\orcidID{0009-0008-8334-0655} \and
Remco Hassing\inst{1}\orcidID{0009-0005-4372-0301} \and
Rob Loves\inst{1}\orcidID{0009-0003-5406-9185} \and
Stephen McCombie\inst{2}\orcidID{0000-0002-6511-9382}}
\authorrunning{G M. Makrakis et al.}
\institute{Maritime IT Security Research Group, NHL Stenden,
Emmen, The Netherlands\\
\email{\{george.makrakis,jeroen.pijpker, remco.hassing, rob.loves\}@nhlstenden.com} \and
Maritime IT Security Research Group, NHL Stenden,
Leeuwarden, The Netherlands\\
\email{stephen.mccombie@nhlstenden.com} 
\and
Department of Computer Science,  University of Groningen,
Groningen, The Netherlands\\
\email{j.pijpker@rug.nl}
}

\maketitle              

\begin{center}
    \centering{\textbf{Accepted Manuscript}}\\
This version of the article has been accepted for publication in CyberICPS 2025 workshop. The final version is available online at SpringerLink.
\end{center}

\begin{abstract}

Cyber threats against the maritime industry have increased notably in recent years, highlighting the need for innovative cybersecurity approaches. Ships, as critical assets, possess highly specialized and interconnected network infrastructures, where their legacy systems and operational constraints further exacerbate their vulnerability to cyberattacks. To better understand this evolving threat landscape, we propose the use of cyber-deception techniques and in particular honeynets, as a means to gather valuable insights into ongoing attack campaigns targeting the maritime sector.

In this paper we present Salty Seagull, a honeynet conceived to simulate a VSAT system for ships. This environment mimics the operations of a functional VSAT system onboard and, at the same time, enables a user to interact with it through a Web dashboard and a CLI environment. Furthermore, based on existing vulnerabilities, we purposefully integrate them into our system to increase attacker engagement. We exposed our honeynet for 30 days to the Internet to assess its capability and measured the received interaction. Results show that while numerous generic attacks have been attempted, only one curious attacker with knowledge of the nature of the system and its vulnerabilities managed to access it, without however exploring its full potential \footnote{This is the accepted manuscript version. The final published version is available at https://doi.org/10.1007/978-3-032-16089-8\_23}.
\end{abstract}

\section{Introduction}\label{sec:intro}	

As one of the most integral components of modern maritime, vessels are equipped with a broad array of cyber systems that encompass both Information Technology (IT) and Operational Technology (OT), distributed across various operational domains within the ship. These include communication systems, propulsion and machinery control, power management, navigation, and cargo handling systems. Designed to operate with minimal human intervention, these systems aim to optimize vessel performance while ensuring safety and reliability. However, their increasing complexity and interconnectivity introduce a broad attack surface. Many of these systems depend on interconnected devices and digital services which, if compromised, could significantly disrupt vessel functionality and operational continuity \cite{dragos2023,mdpi2022}.

Previous studies have demonstrated that Very Small Aperture Terminals (VSATs), satellite-based communication systems commonly deployed to provide Internet connectivity to ships at sea, can be exploited through various attack vectors \cite{pavur2020tale,willbold2024vsaster}. Given their exposure and relatively accessible attack surface, VSAT systems are often regarded as a low-hanging fruit for adversaries seeking initial access. Once compromised, these systems may serve as a pivot point for further reconnaissance, services disruption, or intrusion into more critical onboard systems, thus posing a significant risk to maritime cybersecurity \cite{cydome_labdookhtegan_2025}.


To observe and analyze the behavior of adversaries attempting to exploit digital systems, honeypots and honeynets can be strategically deployed. These cyber-deception systems simulate vulnerable targets, luring attackers into interacting with what they perceive to be legitimate services or devices. During these interactions, valuable insights can be gathered regarding the attacker’s movements and techniques. While honeypots are commonly used in traditional IT environments to simulate services (e.g. SSH, HTTP, or FTP), their deployment in Cyber-Physical Systems (CPS), such as those found aboard maritime vessels, presents unique challenges. These challenges stem from the domain-specific nature of the systems involved, which often include proprietary communication protocols and interactions with physical processes.


In this paper, we propose the design and implementation of a VSAT honeynet capable of attracting, recording, and analyzing early-stage interactions from potential attackers targeting shipboard communication systems. To ensure the system's effectiveness and realism, our design is informed by the ICSvertase framework \cite{kempinski_icsvertase_2023}, while the implementation is grounded in the characteristics of real-world, Internet-exposed VSAT devices. To enhance the credibility of the simulation, the honeynet integrates a Web dashboard and Command Line Interface (CLI), alongside replayed voyage data, thereby mimicking the operational features of an actual maritime communication system. Stemmed from real-world VSAT vulnerabilities, we purposefully integrate them into our system to increase the chances of attacker interaction.


Based on the specific nature of the system we try to simulate, we aim to answer the following Research Questions (RQs):


\begin{itemize}
    \item RQ1: Were there any actors exploiting specific vulnerabilities related to VSAT environments?
    \item RQ2: What were their interactions with Web dashboard and CLI environments?
    \item RQ3: Were there any persistence mechanisms attempted?
\end{itemize}



The collected results from 30 days of deployment indicate that despite the multiple attempts to access the system with generic exploits, only one attacker with knowledge of the nature of the system and its vulnerabilities managed to access it, and perform some familiarization actions inside the environment. Thus, we deduce that specific knowledge is required to meaningfully exploit such systems and gain access to the inner workings of a ship's network. The anonymized raw data to reach to those conclusions are provided in \footnote[1]{https://doi.org/10.5281/zenodo.15469996}.


The rest of the manuscript is organized as follows: preliminary background information is presented in Section \ref{sec:background}, while our the design of our system is detailed in Section \ref{sec:system_design}. The evaluation of Salty Seagull is described in Section \ref{sec:evaluation}, followed by the concluding remarks in Section \ref{sec:conclusions}.

\subsubsection{Ethical Considerations:} The ultimate goal of this research is to improve the security of vessels and their satellite communication systems, by understanding the potential attackers' interactions with them. To this end, all scanning and resources retrieval were conducted via the Shodan and Censys services, without any login attempts or other forms of active interaction with each accessed system.
\section{Background}\label{sec:background}	
In the following, we briefly present the some background information useful to understand the remainder of the paper.

\subsection{Honeypots and Honeynets}

Honeypots are defined by Spitzner as a ``security resource whose value lies in being probed, attacked, or compromised'' \cite{spitzner2002honeypots}. Conceptually, they function as deliberately vulnerable systems that mimic real services or data to deceive and attract malicious actors. Their core objectives are to divert attackers from critical assets, gather detailed intelligence on adversarial behavior, and engage intruders long enough to monitor and analyze their actions. The origins of honeypots trace back to early deception systems described in Cliff Stoll's \textit{The Cuckoo's Egg} \cite{stoll2024cuckoo} and Bill Cheswick's \textit{An Evening With Berferd} \cite{cheswick1992evening}, where attackers were lured into controlled environments to better understand their behavior. These foundational works inspired the Honeynet Project \cite{spitzner2003honeynet}, which advanced the use of honeypots as tools for studying and defending against cyber threats. When multiple honeypots are deployed and interconnected, they form a honeynet—an architecture that has since evolved into a powerful method for capturing adversarial Tactics, Techniques, and Procedures (TTPs) \cite{mitre2025attack}, enhancing both defensive strategies and cyber threat intelligence capabilities.

\subsection{IoT, IIoT and CPS Honeypots}
A plethora of honeypots has been developed and used in the past to acquire information about their respective environments \cite{mahmoud2019deploying,koniaris2013analysis,jiang2007out,srinivasa2022deceptive}. Well known examples of tools used to create honeypots include Cowrie \cite{cowrie}, Conpot \cite{rist2013conpot}, Glastopf \cite{rist2010glastopf}, Dionaea \cite{dionaea}, and T-Pot \cite{telecom2025tpot}. Honeypots have gained considerable attention in industrial environments due to their potential to attract adversaries, such as ransomware gangs, who target these systems for financial gain or adversaries that aim to to inflict denial, manipulation and loss of control, view, and safety \cite{hilt2020caught}. 

Franco et al. compiled a survey on honeypots deployed in the IoT, industrial IoT, CPS and Industrial Control Systems (ICS). They categorized the honeypots based on a multitude of parameters such as levels of interactions, roles, and scalability \cite{franco_survey_2021}. According to this research a significant challenge in deploying honeypots in such environments arises from their low-interaction nature. Most implementations rely on simplified versions of the targeted systems, often lacking the complex functionality and specific vulnerabilities that would typically exist in fully operational systems. This makes these honeypots relatively easy for attackers to identify, as they do not replicate the nuanced behavior of genuine industrial environments. Consequently, the lack of sophisticated interaction can limit the effectiveness of these honeypots in evading detection and capturing valuable intelligence.


\subsection{Shipboard Systems and VSAT}

In the context of CPS, consider the modern ship, which is equipped with a variety of integrated systems designed to enhance operational efficiency and support the crew. These systems encompass a wide range of functionalities, from autopilot controls to sensors monitoring environmental factors, such as water temperature. An example of shipboard systems can be found in Table \ref{tab:shipboard_examples}.





\begin{table}
  \scriptsize
  \centering
  \caption{Examples of Shipboard Systems according to \cite{rajaram2022guidelines}.}
  \label{tab:shipboard_examples}
  \begin{tabular}{ll}
    \toprule
    \multicolumn{1}{c}{\textbf{System}} & \multicolumn{1}{c}{\textbf{Components}} \\
    \midrule
     Communication & \makecell{Satellite Communication System (SATCOM)\\Integrated Communication System (ICS)\\Wireless Local Area Network (WLAN)}\\
    \midrule
     \makecell{Propulsion, Machinery,\\ and Power Control} & \makecell{Engine Governor System\\Fuel Oil System\\Alarm Monitoring \& Control System\\Power Management System\\Emergency Generators and Batteries}\\
     \midrule
     Navigation & \makecell{Electronic Chart Display and Information System (ECDIS)\\Radio Detection and Ranging (RADAR)\\Automatic Identification System (AIS)\\Global Positioning System (GPS)\\Dynamic Positioning System (DPS)\\Global Maritime Distress and Safety System (GMDSS)\\Voyage Data Recorder (VDR)\\Integrated Navigation System (INS)}\\
     \midrule
     Cargo Management & \makecell{Cargo Control Room (CCR)\\Ballast Water System (BWS)}\\
  \bottomrule
\end{tabular}
\end{table}

Among the various systems in a ship, the VSAT is particularly relevant to this study. VSAT is part of SATCOM that enables vessels to maintain internet and television connectivity while at sea. Another use case is the transmission of fishing yield data to cloud services from fishing vessels \cite{rivieramm_vsat}. It operates as a two-way satellite ground station, utilizing a dish antenna connected to a gateway that facilitates a Wide Area Network (WAN). VSATs commonly include Web as well as command line interfaces for configuration and maintenance tasks. Despite their importance in maintaining communications, VSAT systems have previously been found vulnerable to cyber threats \cite{CVE20185266,CVE20185267}. Exploiting these vulnerabilities could provide adversaries with a potential entry point to breach the internal network of the vessel.




\section{System Design}\label{sec:system_design}

In this section, we first introduce the threat model we operate on, then the main design considerations for our honeynet, and finally the technical aspects of the exposed services and the configuration of the environment.

\subsection{Threat Model}

We assume that malicious actors targeting VSAT systems aim to explore these systems and potentially establish a foothold within the vessel's network. These actors are likely to possess knowledge about the use of VSATs on ships and may have access to documentation that details their operation, which could be either publicly or privately available.

The primary objectives of these attackers include either disabling the VSAT system to create confusion for the crew and passengers, and/or exploiting the system to facilitate lateral movement within the ship's internal network. To achieve these goals, attackers would typically leverage both documented and undocumented vulnerabilities, often after gathering basic information, such as version details or build numbers, through manual or automated scanning techniques. Additionally, we assume that attackers will be able to fingerprint the IP address of the VSAT system to ensure that they are targeting a legitimate system of a ship.

\subsection{Design Considerations}

With regards to the defined threat model, we leveraged the ICSvertase framework \cite{kempinski_icsvertase_2023} to design our honeynet. Although our system is not strictly an ICS honeynet, we utilized the framework to guide our design, aligning it with components from \textit{MITRE ATT\&CK\textsuperscript{\textregistered} for ICS} and \textit{MITRE Engage\textsuperscript{TM}}. The main key considerations is what adversary behaviors should a honeynet be designed to capture, and the effective capture of them, while at the same time recognize the importance of capturing such behavior as well as, the ways to incentivize an adversary to exhibit such behaviors.

Thus according to the Engage approaches, we aim to \textit{Collect} adversary tools, observe tactics, and other raw intelligence about the adversary’s activity, \textit{Detect} adversary activity throughout an environment, \textit{Direct} them towards an intended path and, \textit{Reassure} them that the access to an environment is real by adding authenticity to deceptive components. The mapping of such activities and techniques used are shown in Tables \ref{tab:engage} and \ref{tab:attack}.

\begin{table}
  \scriptsize
  \centering
  \caption{Engage Approaches Activities According to ICSvertase Requirements Analysis.}
  \label{tab:engage}
  \begin{tabular}{ll}
    \toprule
    \multicolumn{1}{c}{\textbf{Approach}} & \multicolumn{1}{c}{\textbf{Activity}}\\
    \midrule
    Collect & System Activity Monitoring (EAC0003) \\
     & Software Manipulation (EAC0014)\\
    \midrule
     Detect & Introduced Vulnerabilities (EAC0023)\\
     & Network Analysis \\
     \midrule
     Direct & Introduced Vulnerabilities (EAC0023)\\
     \midrule
     Reassure & Information Manipulation (EAC0015)\\
  \bottomrule
\end{tabular}
\end{table}

\begin{table}
  \scriptsize
  \centering
  \caption{ATT\&CK Techniques According to ICSvertase Requirements Analysis.}
  \label{tab:attack}
  \begin{tabularx}{\linewidth}{llX}
    \toprule
    \multicolumn{1}{c}{\textbf{Technique}} & \multicolumn{1}{c}{\textbf{Functional Feature}} & \multicolumn{1}{c}{\textbf{Data Component}}\\
    \midrule
    Command-Line Interface (T0807) & \makecell{Size : single\\ ICS component : Protocols\\ ICS component : OS\\Logging : file system\\Logging : processes} & \makecell{Application Log Content\\Command Execution\\Process Creation} \\
    \midrule
     Commonly Used Port (T0885) & \makecell{Size : single\\ICS component : OS \\Logging : processes} & \makecell{Network Traffic Content\\Network Traffic Flow}\\
     \midrule
     Default Credentials (T0812) & \makecell{Size : single\\ ICS component : Protocols} & \makecell{Logon Session Creation\\Network Traffic Content} \\
    \midrule
     Device Restart/Shutdown (T0816) & \makecell{Size : single\\ICS component : Runtime}  & \makecell{Application Log Content\\Network Traffic Content\\Network Traffic Flow\\Device Alarm}\\
     \midrule
     \makecell{Exploit Public-Facing Application \\(T0819)} & \makecell{Size : single\\ICS component : Protocols\\ICS component : Runtime} & \makecell{Application Log Content\\Network Traffic Content}\\
     \midrule
     System Firmware (T0857) & \makecell{Size : single\\ICS component : Runtime\\ICS component : Bootloader\\ Logging : file system} & \makecell{Application Log Content\\Firmware Modification\\Network Traffic Content\\Device Alarm}\\
     \midrule
     Remote System Discovery (T0846) & \makecell{Size : single\\ICS component : OS\\Logging : processes} & \makecell{File Access\\Network Traffic Content\\Network Traffic Flow\\Process Creation}\\
     \midrule
     \makecell{Remote System Information \\Discovery (T0888)}  & \makecell{Size : single\\ICS component : OS\\Logging : processes} & \makecell{File Access\\Network Traffic Content\\Network Traffic Flow\\Process Creation}\\
     \midrule
     Valid Accounts (T0859)  & \makecell{Size : single\\ICS component : Protocols} & \makecell{Logon Session Creation\\Logon Session Metadata\\User Account Authentication}\\
  \bottomrule
\end{tabularx}
\end{table}


Collection of activity logs  can reveal adversary activity (EAC0003), while making changes to a system’s software properties and functions can achieve a desired effect (reveal deceptive artifacts and systems) (EAC0014). At the same time introduction of vulnerabilities will motivate the adversary to target specific resources (EAC0023) and the concealment and reveal of both facts and fictions will support a deception story (EAC0015).

The corresponding techniques include exploiting CLIs (T0807) for executing commands, using commonly used ports (T0885) to blend in with normal traffic, and leveraging default credentials (T0812) to gain unauthorized access. Adversaries may also restart or shut down devices (T0816) to disrupt operations, exploit vulnerabilities in public-facing applications (T0819) to gain access, and modify system firmware (T0857) for persistence. Finally, the identified techniques involve discovering remote systems (T0846) and gathering information about them (T0888) to facilitate further attacks, as well as using valid accounts (T0859) to maintain access and evade detection.

Based on the above, the design we propose relates to the traditional classification of \emph{medium-interactive honeynets}. The design choices are made based on the assumption that we can acquire more precise results about techniques attackers might exploit to gain initial foothold to the systems of a vessel. We implement the Web interface of a \emph{Sea Tel VSAT management portal} up to a point where a potential attacker can acquire information about the system and to make changes that however do not have any real impact on the system e.g., upload new firmware. The same applies to its accompanied CLI where a variety of commands can be issued with up to two arguments/options following them. While in this study we explore this specific system, we argue that VSAT systems from other vendors might also include similar vulnerabilities that allow an attacker to gain access to a ship's networks.

Given that the vulnerabilities associated with the studied system (CVE-2018-5267, CVE-2018-5266, CVE-2018-5071, CVE-2018-5728) include unauthenticated access to sensitive resources, we deliberately chose not to implement those that could be trivially exploited by automated bots or Web crawlers. Instead, we selectively implemented vulnerabilities related to the use of default credentials and exposure of sensitive information, which are more likely to attract targeted interest from actors with a specific focus on maritime systems and equipment. Importantly, all incoming connections to our system are logged. Therefore, when attempts are made to exploit any of the referenced CVEs, we interpret this as an indication that the adversary has conducted some degree of reconnaissance or prior research upon identifying our system as potentially vulnerable.
The honeynet is composed of three parts:
\begin{itemize}
    \item The Web service. The service responsible for the front-end and the back-end services of a Sea Tel VSAT.
    \item The Telnet service. The service that provides a CLI interface used to manage the VSAT.
    \item The VDRPlayer service. The service that enables the replay of voyage information to enhance deception.
\end{itemize}

\begin{figure}[!t]
\centering
\includegraphics[width=0.63\textwidth]{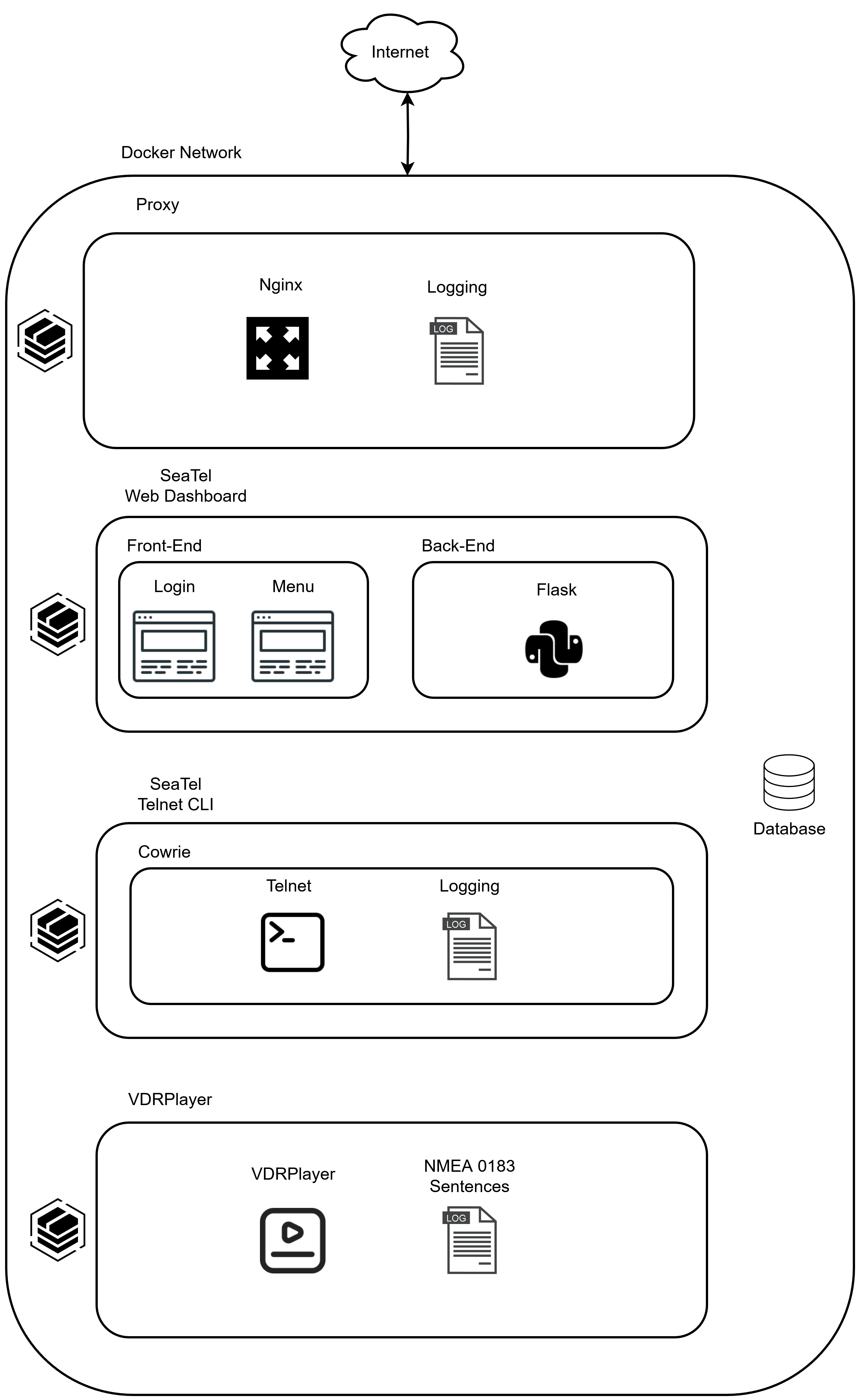}
\caption{The architecture of the proposed honeynet.} 
\label{fig:architecture}
\end{figure}

Our system's architecture is shown in Figure \ref{fig:architecture}. Each of the components incorporates a robust logging system to record attacker trace for further analysis and investigation, based on the aforementioned \textit{MITRE ATT\&CK\textsuperscript{\textregistered} for ICS} Techniques (provided inside brackets in the following subsections).

\subsection{Web Service}

The Web service consists of the proxy, the front-end service and the back-end
service of the Web management portal of a Sea Tel VSAT. We describe each as follows:

\subsubsection{Proxy}

A minimal configuration of the widely used Nginx proxy is deployed to serve as the entry point to the Web service. This setup allows potential adversaries to perform service discovery [T0846] and attempt to exploit a public-facing application [T0819], in alignment with known adversarial tactics. Nginx was selected due to its built-in support for detailed HTTP request and response logging in JSON format, which facilitates effective data analysis. To enhance the realism of the emulated system, we modified the HTTP response headers to mimic those typically returned by a Sea Tel VSAT.

\subsubsection{Front-end}

The front-end of the system comprises static assets, such as HTML, CSS, and JavaScript files, that are served by the back-end component. The user is first presented with a login page, which acts as the landing interface. Upon successful authentication, the appropriate menu page is displayed based on the user’s role. The system defines three distinct user roles: ``User'', ``SysAdmin'', and ``Dealer'', each associated with specific access privileges. Consequently, all front-end resources must be served conditionally [T0859].

\begin{figure}[!t]
\centering
\includegraphics[width=0.7\textwidth]{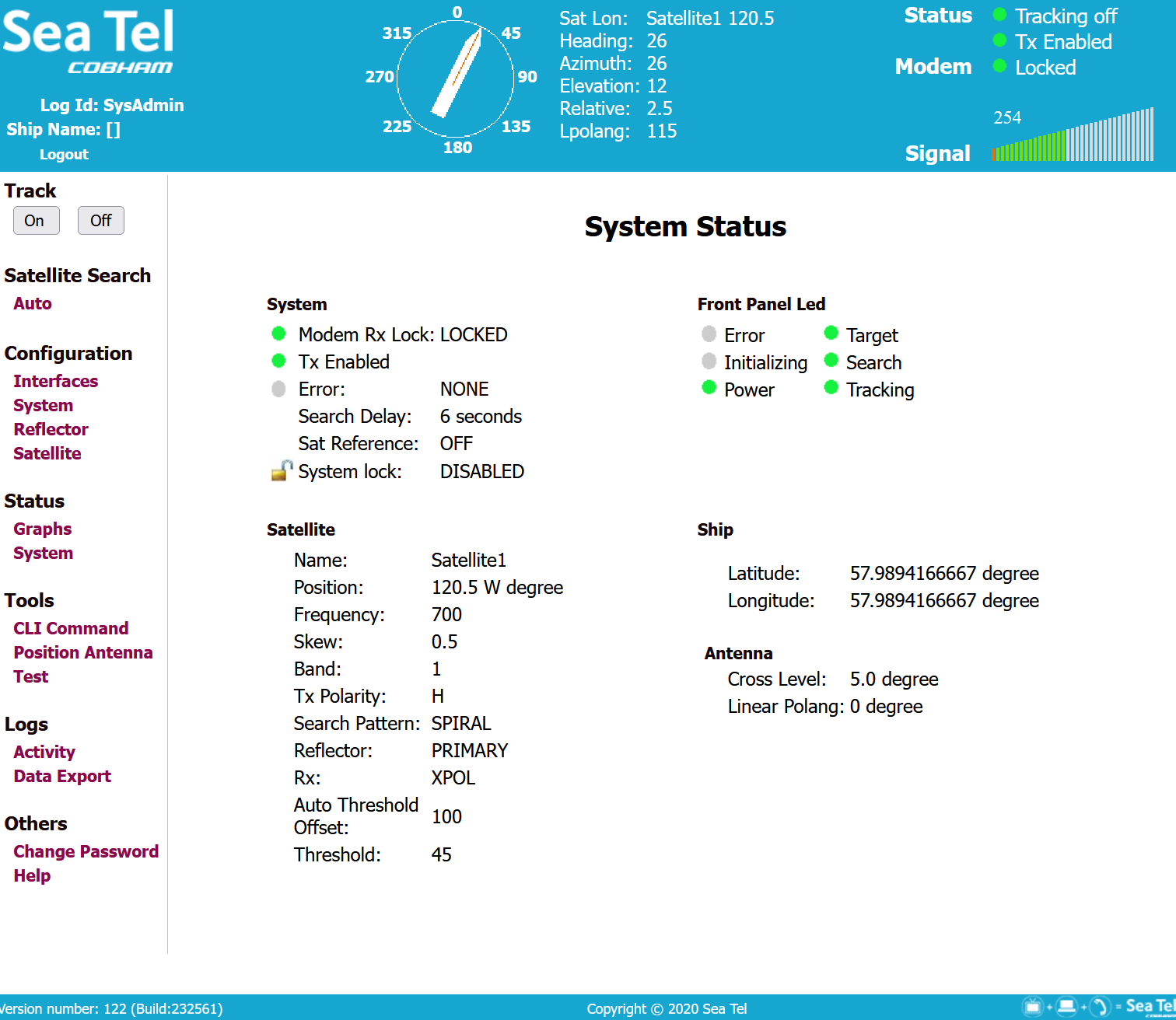}
\caption{An example of the VSAT menu page for the SysAdmin.} 
\label{fig:sysadmin_menu}
\end{figure}

For instance, a user with the ``User'' role is primarily permitted to view operational information related to the satellite and antenna. The ``SysAdmin'' role provides access to system configuration options and diagnostic functionalities, while the ``Dealer'' role grants administrative privileges, including system commissioning and firmware updates. An illustrative example of the dashboard accessible to a ``SysAdmin'' user is presented in Figure \ref{fig:sysadmin_menu}.

Those static files were cloned from other internet exposed VSATs as indicated in \cite{brouwer2024honeyship}. Notice, that some of them were needed to be acquired manually so it will not affect the operation of such exposed systems.

\subsubsection{Back-end}

We implemented the back-end of the Web service in Python Flask. Since the majority of the front-end requests to the back-end are performed via the included JavaScript files, we deduced the data and format that the responses of such requests should have. To this end, we implemented the required endpoints to return all the necessary information based on a combination of realistic and random data. Information related to the nature of the ship such as the heading and coordinates, are drawn from the data replayed from the VDRPlayer Service (see section \ref{subsec:vdrplayer}), while other, such as satellite position and antenna azimuth, are randomly created. At the same time, any changes made via the front-end, are processed and stored in a SQLite database to support the deception narrative.

To avoid the access of our Web-page from bots and spiders, Flask's built-in authentication and authorization mechanisms were used to access each page. The default credentials [T0812] and roles for each user of the system, are stored in the SQLite database. Those credentials can be changed for each user, by accessing the respective page in the Web UI. File uploads for firmware [T0857] and configuration modifications, are saved to explore the potential of an attacker trying to modify the system and maintaining persistence.

Regarding the integrated CLI menu option [T0807], since the access to the vendor's commands manual for the Web interface was not accessible anymore in the public domain, we implemented only a subset of the commands that are found when accessing the system via the Telnet service.

\subsection{Telnet Service}
\label{subsec:Telnet}

The Sea Tel VSAT system features a CLI for management, which operates on its Antenna Control Unit (ACU) and can be accessed via Telnet [T0807]. Although, not exposed by default, we implement it as such to give the impression of a misconfigured setup. To emulate this functionality, we utilize the widely adopted and academically recognized Cowrie SSH and Telnet honeypot \cite{cowrie}. However, since the VSAT CLI interface includes a distinct set of commands and error messages, it is necessary to customize the default Cowrie setup to better align with the real system \cite{seatel_CLI}.

To achieve this, we disabled SSH and implemented a subset of the VSAT CLI commands in Telnet, supporting up to two options for each command. An exception to this rule are commands that were further developed to enhance deception by displaying data from the VDRPlayer service or restarting the system [T0816], as detailed in Section \ref{subsec:vdrplayer}. The default credentials used in the Web service are also applied to the Telnet service, ensuring consistency. If credentials are modified in one place, the change propagates due to the shared SQLite database. Furthermore, to improve realism, fake but plausible parameters such as MAC addresses and ship names are incorporated. All interactions with the Telnet service are logged in JSON format for subsequent analysis.

\subsection{VDRPlayer Service}
\label{subsec:vdrplayer}

The Voyage Data Recorder (VDR) is an important component of any vessel as it collects and provides forensic evidence in the case of an incident/accident, and monitors system performance. A typical VDR system consists of an electronics unit that gathers data from GPS, heading and speed system, ECDIS, AIS, RADAR, voice feeds from bridge, and rudder response. \cite{iacs_vdr}. Commonly it includes two hard drives that mirror each other for redundancy reasons.

By utilizing the open-source VDRPlayer tool \cite{vdrplayer}, we replay such data to the network and feeding them to the back-end and Telnet services. The data are encoded as NMEA 0183 sentences \cite{raymond2019nmea}, that the Web and Telnet services decode and process the data accordingly. This ensures that an attacker with access to both services receives the same data, making it believable that the system belongs to a genuine vessel. When the feed of the data ends, we repeat the replay to keep a constant feed of data to our honeynet.

\subsection{Containerization}
\label{subsec:containerization}

Each of the aforementioned services is deployed as a Docker container on a rented Virtual Private Server (VPS) from a commercial cloud provider. The deployment is managed using separate docker-compose.yml configuration files, one for the Web service and another for the Telnet service, with the VDRPlayer included in the Web service configuration. These files define how the containers are launched, how logs are stored, and how the services interact within the honeynet.

An internal network is established to create the environment, enabling communication between the back-end, Telnet, and VDRPlayer services. For instance, the VDRPlayer replays NMEA data via UDP, which is consumed by both the back-end and Telnet services. The only externally exposed ports are port 80 for the Nginx proxy and port 23 for the Telnet service (T0885), minimizing the attack surface while preserving the honeynet's intended functionality.

\section{Evaluation}\label{sec:evaluation}
In this section, we present the experimental setup, analysis of the
results from a the deployment of our honeynet for 1 month, and some insights from the identified attacks.

\subsection{Experimental Setup}

To evaluate our system, we deployed it on a commercial VPS by a cloud provider located in Europe. The honeynet was hosted on a virtual machine equipped with 4GB of RAM and two virtual CPU cores, running Ubuntu 24.04 as the operating system. To emulate the behavior of a VSAT system, only ports 80 (HTTP) and 23 (Telnet) were left open and accessible from the Internet.

Previous studies in the domain of IT, and IoT/IIoT, CPS and ICS honeynets have explored the deployment of multiple instances distributed across various geographic regions \cite{tambe2019detection}. In contrast, our objective is to realistically simulate the environment of a single vessel. Each ship typically has unique attributes, such as name, call sign, and location, making a single-instance deployment more appropriate for deception. This approach increases the likelihood that adversaries perceive our VSAT system as part of a genuine maritime asset.

To further enhance realism and reduce detectability, we channeled our exposed services through IP address blocks leased from a commercial provider offering both IPv4 and IPv6 addresses. This strategy mitigates the risk of our system being flagged as a honeynet based on a reverse lookup or reputation check of the cloud provider’s IP space. Specifically, we employed Generic Routing Encapsulation (GRE) to tunnel incoming traffic from the leased IP address to the public IP address of the VPS. Subsequently, internal Linux-based routing forwards the traffic to and from the internal Docker network hosting the containerized honeynet.

\subsection{Data Analysis}

\begin{figure}[!h]
\centering
\includegraphics[width=\textwidth]{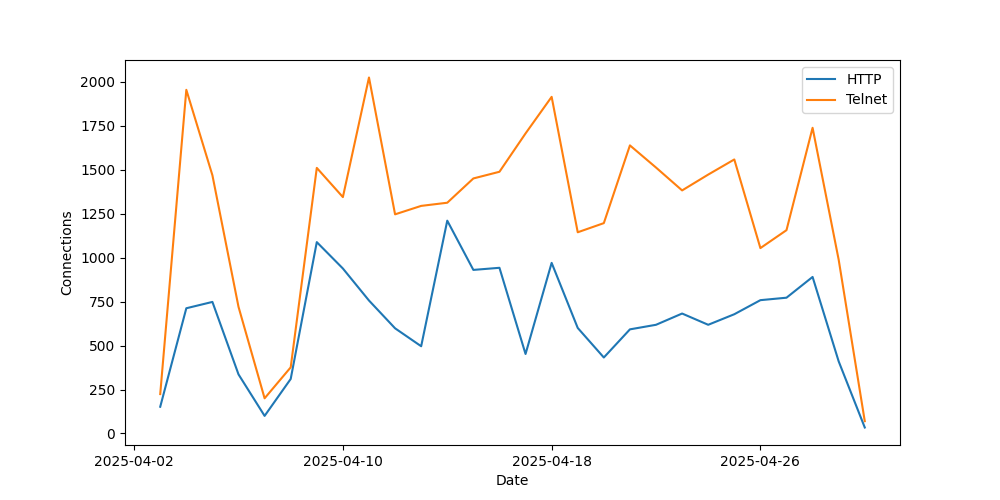}
\caption{Change in the number of log entries gathered per day.} 
\label{fig:logs_http_telnet}
\end{figure}

This setup has been running from Apr 3, 2025 to May 3, 2025. In this period, we have gathered around 16 MB logs from the Web service and 22 MB from the Cowrie honeypot, making a total of 175,290 entries. Those entries include attempts to access our honeynet services from 9,054 distinct IP addresses. Table \ref{tab:geolocations} shows the distribution of Web and Telnet connections based on geographical location. In Figure \ref{fig:logs_http_telnet}, we see the change in log recorded for the two services per day. It is evident that the Telnet service received more attempts than the Web one.

\begin{table}
  \scriptsize
  \centering
  \caption{Top 20 connections received by geolocation.}
  \label{tab:geolocations}
  \begin{tabular}{ll|ll}
    \toprule
    \textbf{Geolocation 1/2} & \textbf{Times} & \textbf{Geolocation 2/2} & \textbf{Times}\\
    \midrule
     China &	31,895 & Indonesia &	1,397 \\
     India &	12,570 & United Kingdom & 1,210\\  
     USA & 6,080 & Iran & 1,215 \\
     The Netherlands & 4,839 &  Singapore & 1,106\\
     Russia & 2,910 & Vietnam & 1,097\\
     Taiwan & 2,150 & Bulgaria &  937\\
     South Korea & 2,070 & Argentina & 937 \\
     Germany & 1,749 & Japan & 858 \\
     Brazil & 1,608 &  Sweden & 818 \\
     Hong Kong &	1,541  & Türkiye & 492 \\
    \bottomrule
    \end{tabular}
\end{table}



\begin{table}
  \scriptsize
  \centering
  \caption{The 10 most used credential combinations in Telnet.}
  \label{tab:credentials}
  \begin{tabular}{lll}
    \toprule
    \textbf{Username} & \textbf{Password} & \textbf{Times}\\
    \midrule
    admin &	1234 & 1,178 \\
    root &	aquario & 1,010 \\
    root &	admin & 962 \\
    root &	root & 686 \\
    root &	(empty) & 670 \\ 
    root &	hi3518 & 666 \\
    admin &	admin & 634 \\
    admin &	password & 632 \\
    ubnt &  ubnt & 630 \\
    admin &	ujMko0admin & 624 \\
    \bottomrule
    \end{tabular}
\end{table}

To answer \textbf{RQ1: Were there any actors exploiting specific vulnerabilities related to VSAT environments?}, we search in our logs for (a) access to the Web and Telnet services via the default credentials (CVE-2018-5266) and (b) direct access to the status of the VSAT via the getSysStatus endpoint (CVE-2018-5728). Until the writing of this paper, only one such attempt has been recorded in the login page of Web dashboard. On the contrary, generic attempts with login credentials, such as those in Table \ref{tab:credentials}, have been logged in the Telnet service. 

The role ``'User'' has been used in one Telnet login attempt, that however had the combination of \emph{``username"=User,password=1001''} and \emph{``username"=User,p-\\assword=User''}, which are invalid in the scope of this honeynet. In the attempt to access the Web dashboard, a single IP address used the credential combinations of \emph{``username=(empty),password=(empty)'', ``username=Dealer,password=seatel-2'', ``username=dealer,password=seatel2'', ``username=user,password=seatel2'' \\and ``username=User,password=seatel1''}. Only the last one is a valid combination providing access to the Menu interface of the Web dashboard. The only records for accessing the \emph{getSysStatus} endpoint indicate that those attempts have been made after the authentication, and not by a direct access.

Among all the requests, we received 19,334 GET and 644 POST actions. Only 8,950 GET and 228 POST requests attempted to access endpoints served by our Web service. Besides those of the single successful login, the majority of the requests revolve mainly around the Login endpoint, which serves the landing page of the VSAT. Many others such as the GET request to \emph{/cgi-bin/iptest.cgi?cmd=iptest.cgi\&-url=\%60wget+http\%3A\%2F\%2F209.200.246\\.240\%3A8081\%2Fmeow\%60\&-time=\%!(NOVERB) HTTP/1.1)}, or the POST one to \emph{/cgi-bin/account\_mgr.cgi}, indicate generic efforts to exploit potentially vulnerable services.


For \textbf{RQ2: What were their interactions with Web dashboard and CLI environments?}, we observe that a number of 196 IP addresses have attempted to access both services. Nevertheless, only logs from the Web dashboard exhibits some form of interaction, after the single successful login recorded. This constitutes of configuring the satellite (\emph{/ConfigSat.html}), setting the antenna parameters (\emph{/cgi-bin/setAntParams}), setting the possition of the ship (\emph{UserShpPosSet.html}) viewing and exporting data from the VSAT ( \emph{/Viewlog.html} and \emph{/DataExport.html}), and attempting to access the Menu for the ``Dealer'' (\emph{/MenuDealerGX.html}). Since there are no successful logins to the Telnet service, there are no meaningful interactions that can be correlated between the two honeynet services.

Regarding \textbf{RQ3: Were there any persistence mechanisms attempted?}, such as change of passwords, configuration or drop of modified firmware, no such attempts were made. There was only one attempt, in the Web service, to escalate privileges to the user ``Dealer'', by attempting to directly access the corresponding Menu page while being authenticated as ``User''. This lack of usage of persistence mechanisms could be attributed to the fact that, besides the change of the password of the user ``User'', all the other endpoints that could potentially enable persistence mechanisms to be established, require the user to first be authenticated with a properly privileged account.

Overall, the above indicate only one attack from a curious party that is somewhat familiar with the particular VSAT system and its vulnerabilities, that is still exploring the nature of such systems. In this attack, the actor was knowledgeable about the user roles and their privileges of the VSAT, but only attempted some basic actions via the account with the lowest permissions. We intend to continue collecting data from our honeynet over the course of the next year. This could enrich the answers to the above RQs, and can provide us with a better understanding of any further adversaries that specifically look for and target VSAT systems.

\subsection{Limitations}
\label{subsec:limitations}

In the current implementation, Salty Seagull is limited to the simulation of a VSAT communication system, with no additional shipboard components integrated. Expanding the honeynet to include other critical subsystems, such as propulsion control and navigation systems, along with vulnerabilities that facilitate lateral movement, could significantly enhance the realism and depth of the deception environment. Nevertheless, the present configuration effectively illustrates how the compromise of an insecure communications system could serve as an initial entry point for adversaries targeting maritime assets.

It is important to note that a sophisticated adversary may attempt to cross-reference the replayed VDR information, such as latitude and longitude coordinates, with external maritime tracking data to assess the authenticity of the simulated vessel. As a potential enhancement, future work could involve the integration of live voyage data from operational commercial vessels. However, acquiring such real-time data introduces substantial challenges, particularly concerning privacy regulations and operational security constraints.

Finally, we acknowledge that the credibility and discoverability of the honeynet could be further strengthened by situating it within a network infrastructure associated with satellite telecommunications. Deploying the system over low Earth orbit satellite networks, such as Starlink, OneWeb, or Amazon’s Project Kuiper, may more convincingly reinforce the perception that the emulated VSAT system belongs to a legitimate maritime platform, thereby improving the effectiveness of the overall deception strategy.
\section{Conclusions}\label{sec:conclusions}

To investigate the threat landscape associated with VSAT systems aboard maritime vessels, we developed a specialized VSAT honeynet designed to attract potential attackers targeting shipboard communication infrastructure. To replicate the operational characteristics of such CPS environments, we implemented both a Web-based dashboard and a CLI, each populated with simulated voyage data. The major findings of this work demonstrate that despite numerous attempts using generic exploits, only a single adversary demonstrated awareness of the system's nature and associated vulnerabilities, successfully gaining access and performing initial reconnaissance within the environment. This suggests that meaningful exploitation of such systems requires specific domain knowledge, underscoring the complexity of compromising a ship's network. Future improvements could include the expansion of the honeynet with more elaborate ship services and components to increase emulation fidelity.




\begin{credits}
\subsubsection{\ackname} We sincerely thank the anonymous reviewers for their insightful comments and valuable suggestions. 
We would like to acknowledge the students participating in the Hack@Sea Minor course in 2024 at NHL Stenden University of Applied Sciences, for their assistance regarding parts of the web development used for this work.

\subsubsection{\discintname}

The authors have no competing interests to declare that
are relevant to the content of this article.

\subsubsection{Data Availability.} The anonymized version of the data collected in this work are made available at \url{https://doi.org/10.5281/zenodo.15469996}.

\end{credits}


%
%
%
\bibliographystyle{splncs04}
\bibliography{references}

@misc{cydome_labdookhtegan_2025, 
    title={Lab Dookhtegan cyber attack on Iranian oil tankers disrupts operations},
    author={{CYDOME}},
    note={\url{https://cydome.io/lab-dookhtegan-cyber-attack-on-iranian-oil-tankers-disrupts-operations/}, accessed 2025-04-27}
}

@misc{dragos2023, 
  title        = {ICS/OT Cybersecurity Considerations for Maritime Transportation},
  author       = {Liz Martin and Blake Benson},
  year         = {2023},
  url          = {https://hub.dragos.com/hubfs/116-Whitepapers/Dragos_WP_ICS_OTCybersecuritMaritimeTransp_Final%20(1).pdf},
  note         = {Accessed: 2025-04-03}
}

@article{mdpi2022,
    AUTHOR = {Akpan, Frank and Bendiab, Gueltoum and Shiaeles, Stavros and Karamperidis, Stavros and Michaloliakos, Michalis},
    TITLE = {Cybersecurity Challenges in the Maritime Sector},
    JOURNAL = {Network},
    VOLUME = {2},
    YEAR = {2022},
    NUMBER = {1},
    PAGES = {123--138},
    URL = {https://www.mdpi.com/2673-8732/2/1/9},
    ISSN = {2673-8732},
    DOI = {10.3390/network2010009}
}

@inproceedings{willbold2024vsaster,
  title={VSAsTer: Uncovering Inherent Security Issues in Current VSAT System Practices},
  author={Willbold, Johannes and Schloegel, Moritz and Bisping, Robin and Strohmeier, Martin and Holz, Thorsten and Lenders, Vincent},
  booktitle={Proceedings of the 17th ACM Conference on Security and Privacy in Wireless and Mobile Networks},
  pages={288--299},
  year={2024}
}

@inproceedings{pavur2020tale,
  title={A tale of sea and sky on the security of maritime VSAT communications},
  author={Pavur, James and Moser, Daniel and Strohmeier, Martin and Lenders, Vincent and Martinovic, Ivan},
  booktitle={2020 IEEE Symposium on Security and Privacy (SP)},
  pages={1384--1400},
  year={2020},
  organization={IEEE}
}

@book{spitzner2002honeypots,
  title={Honeypots: tracking hackers},
  author={Spitzner, Lance},
  year={2002},
  publisher={Addison-Wesley Longman Publishing Co., Inc.}
}

@book{stoll2024cuckoo,
  title={The cuckoo's egg: tracking a spy through the maze of computer espionage},
  author={Stoll, Cliff},
  year={1989},
  publisher={Simon and Schuster}
}

@inproceedings{cheswick1992evening,
  title={An Evening with Berferd in which a cracker is Lured, Endured, and Studied},
  author={Cheswick, Bill},
  booktitle={Proc. Winter USENIX Conference, San Francisco},
  pages={20--24},
  year={1992}
}

@article{spitzner2003honeynet,
  title={The honeynet project: Trapping the hackers},
  author={Spitzner, Lance},
  journal={IEEE Security \& Privacy},
  volume={1},
  number={2},
  pages={15--23},
  year={2003},
  publisher={IEEE}
}

@misc{mitre2025attack, 
    title={{MITRE ATT\&CK®}},
    author={{MITRE}},
    note={\url{https://attack.mitre.org/}, accessed 2025-04-03}
}

@article{franco_survey_2021,
  title={A survey of honeypots and honeynets for internet of things, industrial internet of things, and cyber-physical systems},
  author={Franco, Javier and Aris, Ahmet and Canberk, Berk and Uluagac, A Selcuk},
  journal={IEEE Communications Surveys \& Tutorials},
  volume={23},
  number={4},
  pages={2351--2383},
  year={2021},
  publisher={IEEE}
}

@inproceedings{mahmoud2019deploying,
  title={Deploying a University Honeypot: A case study},
  author={Mahmoud, Rasmi-Vlad and Pedersen, Jens Myrup},
  booktitle={CEUR Workshop Proceedings},
  volume={2443},
  pages={27--38},
  year={2019},
  organization={CEUR Workshop Proceedings}
}

@inproceedings{koniaris2013analysis,
  title={Analysis and visualization of SSH attacks using honeypots},
  author={Koniaris, Ioannis and Papadimitriou, Georgios and Nicopolitidis, Petros},
  booktitle={Eurocon 2013},
  pages={65--72},
  year={2013},
  organization={IEEE}
}

@inproceedings{jiang2007out,
  title={“Out-of-the-box” monitoring of VM-based high-interaction honeypots},
  author={Jiang, Xuxian and Wang, Xinyuan},
  booktitle={International Workshop on Recent Advances in Intrusion Detection},
  pages={198--218},
  year={2007},
  organization={Springer}
}

@inproceedings{srinivasa2022deceptive,
  title={Deceptive directories and “vulnerable” logs: a honeypot study of the LDAP and log4j attack landscape},
  author={Srinivasa, Shreyas and Pedersen, Jens Myrup and Vasilomanolakis, Emmanouil},
  booktitle={2022 IEEE European Symposium on Security and Privacy Workshops (EuroS\&PW)},
  pages={442--447},
  year={2022},
  organization={IEEE}
}

@inproceedings{rajaram2022guidelines,
  title={Guidelines for cyber risk management in shipboard operational technology systems},
  author={Rajaram, Priyanga and Goh, Mark and Zhou, Jianying},
  booktitle={Journal of Physics: Conference Series},
  volume={2311},
  number={1},
  pages={012002},
  year={2022},
  organization={IOP Publishing}
}

@article{hilt2020caught,
  title={Caught in the act: Running a realistic factory honeypot to capture real threats},
  author={Hilt, Stephen and Maggi, Federico and Perine, Charles and Remorin, Lord and R{\"o}sler, Martin and Vosseler, Rainer},
  journal={Trend Micro Research},
  year={2020}
}

@mastersthesis{brouwer2024honeyship,
  author       = {Brouwer, SC},
  title        = {HoneyShip: A Maritime VSAT Honeypot to Collect Cyberattacks and Analyze Threats},
  school       = {Rijksuniversiteit Groningen},
  address      = {9712 CP Groningen, Netherlands},
  year         = {2024},
}

@misc{cowrie, 
    title={Cowrie SSH/telnet honeypot},
    author={Michel Oosterhof},
    note={\url{https://github.com/micheloosterhof/cowrie}, accessed 2025-04-01}
}

@article{rist2010glastopf,
  title={Know your tools: Glastopf-a dynamic, low-interaction web application honeypot},
  author={Rist, Lukas and Vetsch, Sven and Kossin, Marcel and Mauer, Michael},
  journal={The Honeynet Project},
  volume={4},
  pages={2},
  year={2010}
}

@article{rist2013conpot,
  title={Conpot ics/scada honeypot},
  author={Rist, Lukas and Vestergaard, Johnny and Haslinger, Daniel and Pasquale, Andrea and Smith, John},
  journal={Honeynet Project (conpot. org)},
  year={2013}
}

@misc{dionaea, 
    title={Web Honeypot},
    author={Dino Tools},
    note={\url{https://github.com/DinoTools/dionaea/}, accessed 2025-04-01}
}

@article{telecom2025tpot,
  title={T-Pot: A Multi-Honeypot Platform},
  author={Deutsche Telekom AG},
  journal={Honeynet Project},
  year={2025}
}

@misc{vdrplayer,
    title={VDRplayer - Play Voyage Data Recorder files over IP link.},
    author={transmitterdan},
    note={\url{https://github.com/transmitterdan/VDRplayer}, accessed 2025-04-01}
}

@misc{CVE20185266, 
    title={CVE-2018-5266},
    author={MITRE},
    note={\url{https://nvd.nist.gov/vuln/detail/cve-2018-5266}, accessed 2025-04-03}
}

@misc{CVE20185267, 
    title={CVE-2018-5267},
    author={MITRE},
    note={\url{https://nvd.nist.gov/vuln/detail/cve-2018-5267}, accessed 2025-04-03}
}

@misc{iacs_vdr, 
    title={Recommendations on Voyage Data Recorder},
    author={IACS},
    note={\url{https://web.archive.org/web/20230202060115/https://iacs.org.uk/download/1871}, accessed 2025-04-01}
}

@misc{seatel_CLI, 
    title={Document IMA CLI  Protocol Specification},
    author={SeaTel},
    note={\url{https://www.yumpu.com/en/document/read/50984924/document-ima-cli-protocol-specification-livewire-connections-ltd}, accessed 2025-04-03}
}

@inproceedings{tambe2019detection,
  title={Detection of threats to IoT devices using scalable VPN-forwarded honeypots},
  author={Tambe, Amit and Aung, Yan Lin and Sridharan, Ragav and Ochoa, Mart{\'\i}n and Tippenhauer, Nils Ole and Shabtai, Asaf and Elovici, Yuval},
  booktitle={Proceedings of the Ninth ACM Conference on Data and Application Security and Privacy},
  pages={85--96},
  year={2019}
}

@inproceedings{kempinski_icsvertase_2023,
	address = {Benevento Italy},
	title = {{ICSvertase}: {A} {Framework} for {Purpose}-based {Design} and {Classification} of {ICS} {Honeypots}},
	isbn = {979-8-4007-0772-8},
	shorttitle = {{ICSvertase}},
	url = {https://dl.acm.org/doi/10.1145/3600160.3605020},
	doi = {10.1145/3600160.3605020},
	language = {en},
	urldate = {2024-10-17},
	booktitle = {Proceedings of the 18th {International} {Conference} on {Availability}, {Reliability} and {Security}},
	publisher = {ACM},
	author = {Kempinski, Stash and Ichaarine, Shuaib and Sciancalepore, Savio and Zambon, Emmanuele},
	month = aug,
	year = {2023},
	pages = {1--10},
}

@article{raymond2019nmea,
  title={Nmea revealed},
  author={Raymond, Eric S},
  journal={URL https://gpsd. gitlab. io/gpsd/NMEA. html},
  year={2019}
}

@misc{rivieramm_vsat, 
    title={Fishing vessel owners turn to VSAT},
    author={Rivieramm},
    note={\url{https://www.rivieramm.com/opinion/opinion/fishing-vessel-owners-turn-to-vsat-35069}, accessed 2025-04-01}
}
%




\end{document}